\documentclass[11pt,a4paper]{article}
\usepackage{jcappub}
\usepackage{bm}
\usepackage{amsmath}
\def\dd{\mathrm{d}}
\def\mcP{\mathcal{P}}

\def\Mpl{M_{\rm Pl}}

\def\ABG{A_{\rm BG}}

\title{
Does Anisotropic Inflation Produce a Small Statistical Anisotropy?
}
\author[a]{Tomohiro Fujita,}
\author[a]{Ippei Obata,}
\affiliation[a]{Department of Physics, Kyoto University, Kyoto, 606-8502, Japan}
\emailAdd{t.fujita@tap.scphys.kyoto-u.ac.jp}
\emailAdd{obata@tap.scphys.kyoto-u.ac.jp}
\abstract{
Anisotropic inflation is an interesting model %of inflation
 with an U(1) gauge field and it predicts the statistical anisotropy of the curvature perturbation
characterized by a parameter $g_*$. 
However, we find that the background gauge field does not follow the classical attractor solution due to the stochastic effect.
We develop the stochastic formalism of a vector field and solve Langevin and Fokker-Planck equations.
It is shown that this model is excluded by the CMB constraint $g_*\le 10^{-2}$
with a high probability about 
$99.999\%$.
}

\arxivnumber{XXXX.XXXXX}

\usepackage{amsfonts}

\begin{document}

\begin{flushright}
KUNS-2708
\end{flushright}

\maketitle

%************************************************************************************%
%
%
%
%====================================================================================%
\section{Introduction}
%====================================================================================%

There is no room for doubt that we face a precision cosmology era.
 Current observations of cosmic microwave background (CMB) have enabled us to probe an imprint of detailed initial conditions in our universe, namely to refine the predictions of the inflationary universe up to a percent level.
 The theory of inflation elegantly explains the origin of primordial density fluctuations as the curvature perturbation produced during inflation.
 So far, CMB missions have revealed that curvature perturbation is well described by a vacuum fluctuation of inflaton, since its statistical properties are nearly scale-invariant and isotropic.
%, which are sufficiently consistent with observations.

At the same time, however, we also expect that an indication of novel signatures, such as primordial anisotropies in early universe, might be discovered 
%from an upcoming more precised data.
by upcoming observations.
 In fact, there have been many attempts to explore a statistical anisotropy of curvature perturbation both observationally and theoretically, inspired by the appearance of anisotropic inflation \cite{Watanabe:2009ct, Dulaney:2010sq, Dimastrogiovanni:2010sm, Gumrukcuoglu:2010yc, Watanabe:2010fh, Kanno:2010nr, Watanabe:2010bu, Soda:2012zm, Maleknejad:2012fw, Ohashi:2013qba}.
 Anisotropic inflation naturally accommodates the background of U(1) gauge field due to its kinematical coupling to inflaton, which violates the conformal invariance of gauge field and makes fluctuations of gauge quanta grow on super-horizon scales.
 The growing background gauge field eventually back-reacts on the motion of inflaton and the energy density of the gauge field settles in a constant value during inflation. Thus the gauge field has an attractor solution.
 Intriguingly, such a sustained vector field supports the accelerated anisotropic expansion of the universe, and generates a quadrupole anisotropy in the two-point function of the curvature perturbation.
After the proposal of anisotropic inflation, many cosmological predictions and theoretical extensions have been discussed \cite{Murata:2011wv, Karciauskas:2011fp, Dimopoulos:2011ws, Dimopoulos:2011pe, Yamamoto:2012sq, Bartolo:2012sd, Funakoshi:2012ym, Shiraishi:2013vja, Abolhasani:2013zya, Lyth:2013kah, Ohashi:2013mka, Ohashi:2013pca, Ramazanov:2013wea, Shiraishi:2013oqa, Thorsrud:2013kya, Emami:2014tpa, Naruko:2014bxa, Blanco-Pillado:2015dfa, Bartolo:2015dga, Ito:2015sxj, Choi:2015wva, Abolhasani:2015cve, Ito:2016aai, Rostami:2017wiy, Sugiyama:2017ggb, Ito:2017bnn, Holland:2017cza}, the part of which has also been studied in the context of an inflation model with a charged scalar field \cite{Emami:2010rm, Emami:2011yi, Emami:2013bk, Abolhasani:2013bpa, Emami:2015qjl} and the model of solid inflation \cite{Bartolo:2013msa, Akhshik:2014gja, Bartolo:2014xfa}.
 The prediction of anisotropic inflation can be tested by the quadrupole anisotropy in the CMB angular power spectrum, while there has been no evidence of such a signal and it has been constrained smaller than a few percent level~\cite{Kim:2013gka, Ade:2015lrj}.

 In light of the present situation, we revisit the possibility to generate  a small statistical anisotropy allowed by the current observations in  anisotropic inflation.
The current observation data %shows us that it should be smaller than $10^{-2}$,
constraining the statistical anisotropy to be smaller than a few percents
requires that the energy fraction of background gauge field should be sufficiently small.
In that case, in anisotropic inflation, the growth rate of the energy density of the gauge field should also be tiny and
it would take an enormous time for the gauge field energy  to reach the constant attractor value. As we show in this paper, this situation is analogous to
a scalar field with a very flat potential. 
Even though the potential of the scalar field has a minimum,
if the curvature of the potential around the minimum is too small, the  stochastic force from the quantum fluctuation wins the classical  force from the gradient of the potential, and the scalar field is not expected to sit in the minimum.
The classical approach to analyze the background system is no longer valid  and quantum kicks to the background should be taken into account in such cases.
In anisotropic inflation, the same argument applies to the background dynamics of the gauge field.
As a result, the prediction of the primordial anisotropy should be altered by the stochastic effect of the gauge field which does not necessarily follow the attractor solution.

The stochastic formalism of scalar field has been developed to handle the cases where the fluctuations significantly influence the background~\cite{Starobinsky:1986fx, Linde:1986fd, Sasaki:1987gy, Nambu:1988je, Nakao:1988yi, Mollerach:1990zf, Starobinsky:1994bd, Finelli:2008zg, Finelli:2010sh, Enqvist:2011pt, Martin:2011ib, Kawasaki:2012bk, Fujita:2013cna, Fujita:2014tja, Vennin:2015hra, Tokuda:2017fdh}.
This formalism takes into account the effect of quantum fluctuations as a stochastic noise term in the equation of motion for a scalar field averaged over a scale much larger than Hubble scale.
Consequently, the dynamics of the scalar field on a large scale with quantum kicks can be solved in a probabilistic way. For example, one can calculate
the probability distribution function of its field value. 
 In this work, therefore, we develop the stochastic formalism of a vector field in the original model of anisotropic inflation.
We derive the Langevin equation and the Fokker-Planck equation in an analogous way to the scalar field case to calculate the dynamics of the gauge field on a large scale.
 We find that the gauge field fluctuations are piled up and quickly overwhelm the classical attractor solution. As a result, the predicted statistical anisotropy becomes far beyond the observational upper bound. This result is consistent with the previous study in which the deviation from the attractor solution is discussed without the stochastic formalism~\cite{Bartolo:2012sd}.
In addition, we find the probability that the statistical anisotropy of the curvature perturbation is not larger than current upper bound is around 0.001\%, irrespective of the model parameter of anisotropic inflation, if the the gauge field is in the stochastic equilibrium when the CMB modes exit the horizon.
 Equivalently, the prediction of the conventional anisotropic inflation scenario is excluded with about $99.999\%$.
 It concludes that the realization of anisotropic inflation is quite challenging under the current observational situations.

This paper is organized as follows.
In Sec.~\ref{Review on Anisotropic Inflation},
the original anisotropic inflation model and its prediction
of the statistical anisotropy in the curvature power spectrum
are reviewed. In Sec.~\ref{Classical versus Stochastic Motion},
we compare the classical motion of the background and the effect of the fluctuations on it in an intuitive way.
Then we develop the stochastic formalism of a vector field
and derive a Langevin equation in Sec.~\ref{Stochastic Formalism of Vector Field}.
In Sec.~\ref{Probability Distribution of Anisotropy},
we solve a Fokker-Planck equation and obtain the probability
distribution function of the statistical anisotropy
generated in anisotropic inflation.
Sec.~\ref{Summary and Discussion} is devoted to
the summary and discussion of the results of this paper.
In appendices \ref{Derivation of the Langevin Equation} and \ref{Derivation of the Fokker-Planck Equation}, the derivations of the Langevin and Fokker-Planck
equations are presented.

%====================================================================================%
\section{Review on Anisotropic Inflation}
\label{Review on Anisotropic Inflation}
%====================================================================================%

In this section, we present a short review of the anisotropic inflation model.
%====================================================================================%
%\subsection{Background dynamics}
%====================================================================================%
In this model, the inflaton $\varphi$ is coupled to an Abelian gauge field $F_{\mu\nu} = \partial_\mu A_\nu - \partial_\nu A_\mu$.
The action is given by
\begin{align}
S %&= S_{\text{EH}} + S_{\text{inflaton}} + S_{\text{gauge}} \notag \\
   &=\int dx^4\sqrt{-g}\left[\dfrac{1}{2}\Mpl^2R-\dfrac{1}{2}(\partial_\mu\varphi)^2 -V(\varphi) - \dfrac{1}{4}f(\varphi)^2F_{\mu\nu}F^{\mu\nu} \right]  \label{eq:action} \,,
\end{align}
where $\Mpl$ is the reduced Planck mass, $R$ is Ricci scalar, and $V(\varphi)$ is the inflaton potential.
For the gauge field, we choose the radiation gauge as $A_0 = 0, \ \partial_i A_i = 0$ and assume that it has a time-dependent background value 
\begin{equation}
\bm{A}(t)=\ABG(t)\, \bm{e}_x
\end{equation}
with a fixed direction. Here, its direction is taken to be x-direction $\bm{e}_x$ without loss of generality.
Based on this setup we can generally adopt the metric of Bianchi
type I spacetime which has the y-z rotational symmetry:
% due to the existence of x-directed background vector variable
\begin{equation}
ds^2=-dt^2+e^{2\alpha(t)}\left[ e^{ - 4\sigma(t)}dx^2 + e^{2\sigma(t)}(dy^2 + dz^2) \right] \equiv -dt^2 + a(t)^2 dx^2 + b(t)^2(dy^2 + dz^2) \ .
\end{equation}
$e^\alpha$ is an isotropic scale factor and $\sigma$ represents a deviation from the isotropic expansion. They satisfy
\begin{equation}
3M_p^2H^2 = 3M_p^2\dot{\sigma}^2 + \dfrac{1}{2}\dot{\bar{\varphi}}^2 + V(\bar{\varphi}) + \rho_E,\qquad
\ddot{\sigma} = -3H\dot{\sigma} + \dfrac{2}{3\Mpl^2}\rho_E,
\label{eq: ani}
\end{equation}
where dot denotes time derivative $\partial_t$, $H \equiv \dot{\alpha}$ is the Hubble parameter, and we defined the energy density of the background electric field as
\begin{equation}
\rho_E = \dfrac{1}{2}\bm{E}^2 \equiv \dfrac{1}{2}\dfrac{f^2\dot{A}^2_{\rm BG}}{a^2} \label{eq: rho0} \ .
\end{equation}
The equations of motion (EoMs) for the background inflaton $\bar{\varphi}(t)$ and the gauge field $\ABG(t)$ are obtained as
\begin{align}
\ddot{\bar{\varphi}} +3H\dot{\bar{\varphi}} + \partial_{\bar{\varphi}}V = 2\dfrac{\partial_{\bar{\varphi}}f}{f}\rho_E \,,\qquad
\left(f^2\dot{A}_{\rm BG}e^{\alpha + 4\sigma}\right)^{\text{\large$\cdot$}} = 0 \label{eq: v} \ ,
\end{align}
These equations can be solved by adopting an ansatz for the gauge kinetic function $f(\varphi)$,
\begin{equation}
f(\varphi) = \exp\left[ \dfrac{2c}{\Mpl^2}\int\dfrac{V}{\partial_{\varphi}V}\dd \varphi \right],
\qquad (c>1)
\end{equation}
where $c$ is a constant model parameter which is assumed to be larger than unity in this model. 
Under the slow-roll approximation,
one can find the solutions~\cite{Watanabe:2009ct}
\begin{align}
&f(\varphi)\propto e^{-2\alpha-2\sigma}\sqrt{1+\Omega\, e^{-4(c-1)\alpha+4\sigma}},
\label{f general}
\\
&\rho_E(t) = \frac{\rho_E^{\rm att}}{1+\Omega\, e^{-4(c-1)\alpha+4\sigma}},
\label{rho general}
\\
&\rho_E^{\rm att} \equiv \dfrac{3}{2}\dfrac{c-1}{c^2}\epsilon_V \Mpl^2 H^2,
\label{rho att}
\end{align}
where $\Omega$ is a constant determined by the initial value of $\rho_E(t)$ and $\epsilon_V\equiv \Mpl^2(\partial_{\varphi}V/V)^2/2$.
Note that as the term with $\Omega$ exponentially decays due to the cosmic expansion, $\rho_E(t)$ approaches $\rho_E^{\rm att}$. It indicates this system has an attractor solution. In the attractor regime, the inflaton potential force is balanced with not only the Hubble friction but also the additional friction from the gauge field (i.e. $2(\partial_{\bar{\varphi}}f/f)\rho_E$ term in \eqref{eq: v}), and the terminal velocity is slowed down by the factor of $c$, 
\begin{equation}
\dot{\bar{\varphi}} = -\dfrac{\sqrt{2\epsilon_V}\Mpl H}{c} \,.
\label{eq: infat}
\end{equation}
If $\rho_E$ is smaller (larger) than the attractor value $\rho_E^{\rm att}$, the additional friction is weaker (stronger) and the inflaton velocity becomes faster (slower) which enhances (suppresses) $\rho_E$ through the gauge kinetic function $f(\varphi)$. Therefore the attractor solution is realized in which an anisotropic expansion of the universe is kept with
\begin{equation}
\dfrac{\dot{\sigma}}{H} = \dfrac{c-1}{3c}\epsilon_H \,.
\end{equation}
where $\epsilon_H = \epsilon_V/c$.

%====================================================================================%
%\subsection{Statistical anisotropy of curvature perturbation}
%====================================================================================%

Next, we describe the statistical anisotropy of the curvature perturbation sourced by the gauge field fluctuations in this model.
 Hereafter, for simplicity we neglect $\sigma$ in the metric, because its time variation needs to be sufficiently small for the compatibility with observations, as we see below. 

First, we calculate the evolution of gauge field perturbation $\delta A_i$
which is expanded as
\begin{align}
\delta A_i(\tau, \bm{x}) = \sum_{\lambda = 1,2}\int \dfrac{\dd^3 k}{(2\pi)^3}e^{i\bm{k}\cdot\bm{x}}
e^\lambda_i(\hat{\bm{k}}) \left[ \delta A^\lambda_k(\tau)\,\hat{a}^\lambda_{\bm{k}} + \delta A^{\lambda*}_k(\tau)\, \hat{a}^{\lambda \dagger}_{-\bm{k}} \right], 
\end{align}
where $e^\lambda_i(\hat{\bm{k}})$ is the polarization vector and  
$\hat{a}^\lambda_{\bm{k}},\hat{a}^{\lambda \dagger}_{-\bm{k}}$ are the creation/annihilation
operators satisfying the commutation relation,
$[a^\lambda_{\bm{k}}, \ a^{\lambda' \dagger}_{-\bm{k}'}] = (2\pi)^3\delta^{\lambda \lambda'}\delta(\bm{k} + \bm{k}')$.
Then the EoM for each polarization mode  reads
\begin{equation}
(f\delta A^\lambda_k)'' + \left( k^2 - \dfrac{f''}{f} \right)f\delta A^\lambda_k = 0 \,,
\end{equation}
where prime denotes the derivative with respective to conformal time $\tau\simeq -1/aH$.
 Assuming $f \propto \tau^n$, %and $f''/f = n(n-1)/\tau^2$,
one finds  the solution with the Bunch Davies initial condition,
\begin{equation}\label{dA solution}
f\delta A_k = \dfrac{\sqrt{-\pi \tau}}{2}H^{(1)}_{n - \frac{1}{2}}(-k\tau),
\end{equation}
where $H_\nu^{(1)}(z)$ is the Hankel function of the first kind.
 Note that we omitted the polarization label $\lambda$, since both modes have the same solution.
 Introducing the electric and magnetic component of the gauge field as
\begin{equation}
\delta E_i \equiv -\dfrac{f}{a^2}\delta A_i' \,, \qquad \delta B_i \equiv \dfrac{f}{a^2}\epsilon_{ijk}\partial_j \delta A_k \label{eq: elemag} \,,
\end{equation}
one can compute their behaviors in the super-horizon limit as
\begin{equation} \label{delta Ek}
\delta E_k = i\dfrac{\Gamma(n+\frac{1}{2})}{\sqrt{\pi}}H^2\left(\dfrac{2}{k}\right)^{3/2}\left(\dfrac{2}{-k\tau}\right)^{n-2} \ , \quad \delta B_k = -i\dfrac{\Gamma(n-\frac{1}{2})}{\sqrt{\pi}}H^2\left(\dfrac{2}{k}\right)^{3/2}\left(\dfrac{2}{-k\tau}\right)^{n-3} \ .
\end{equation}
 Therefore, the perturbation of the electric field does not decay on super horizon scales for $n\geq2$.
 Then the electric field is much stronger than the magnetic field on super horizon scales, and we can ignore the magnetic component.

In the attractor regime of  anisotropic inflation, $n = 2$ ~is approximately realized because of $f\propto a^{-2}$ (see \eqref{f general}), and $\delta E_k$ stays constant on super horizon scale. $\delta E_k$ persistently sources the inflaton perturbation $\delta\varphi$ through the gauge kinetic interaction. As a result, the curvature perturbation which is $\zeta \equiv -H\delta\varphi/\dot{\bar{\varphi}}$ at flat slicing acquires an additional contribution from the electric fields on top of the vacuum fluctuation. One can show that its dimensionless power spectra
\begin{equation}
\langle \hat{\zeta}_{\bm{k}} \hat{\zeta}_{\bm{k}'} \rangle \equiv 
(2\pi)^3\delta(\bm{k} + \bm{k}')\, \dfrac{2\pi^2}{k^3}\mcP_\zeta(k)\,,
\end{equation}
from the vacuum fluctuation $\mcP_\zeta^{\rm (0)}$ and sourced by the electric fields $\mcP_\zeta^{\rm E}$ are given by
\begin{equation}
\mcP_\zeta^{(0)}=\dfrac{c H^2}{8\pi^2\Mpl^2\epsilon_H},
\qquad
\mcP_\zeta^{E}=\dfrac{2c^2\rho_E N_k^2}{\pi^2\Mpl^4\epsilon_H^2}\sin^2\theta\,,
\end{equation}
where $\theta$ denotes the angle between the wave vector and the background gauge field, $\cos\theta\equiv \hat{k}\cdot \bm{e}_x$. $N_k\equiv \ln (a_{\rm end} H/k)$ is the e-fold number at the horizon crossing of the mode in interest
where $a_{\rm end}$ is the scale factor at  the end of inflation.
The total power spectrum of the curvature perturbation can be written
with the statistical anisotropy parameter $g_*$~\footnote{Another definition of $g_*$, $\mcP_\zeta(\bm{k})=\mcP_{\zeta}^{(0)}(k)\left( 1 + g_*\cos^2\theta \right)$
is also often seen in the literature. However, we use the definition \eqref{g* expression} following Ref.~\cite{Watanabe:2010fh}. In this case, $g_*$ produced in anisotropic inflation can take a value larger than unity.}
\begin{align}
\mcP_\zeta(\bm{k}) &= \mcP_{\zeta}^{(0)}(k)\left( 1 + g_*\sin^2\theta \right),
\\
\qquad g_* &= \frac{16c\, N_k^2}{\epsilon_H \Mpl^2 H^2}\,\rho_E
\,\xrightarrow{\rho_E \to \rho_E^{\rm att}}\, 
24(c-1)N_k^2,
\label{g* expression}
\end{align}
where we substituted the attractor value of $\rho_E$ \eqref{rho att}
in \eqref{g* expression}.
The non-detection of the statistical anisotropy by the CMB observations puts severe constraints on $g_*$ and consequently on the model parameter $c$,
\begin{equation}
|g_*| \lesssim 10^{-2} \quad\Longrightarrow\quad
c-1 \lesssim 10^{-7}\left(\frac{N_{\rm CMB}}{60}\right)^2,
\end{equation}
where $N_{\rm CMB}$ denotes the e-folds at which the CMB scale modes
exit the horizon.
It should be stressed that we obtain the constraint on $c-1$ because
the background electric energy density $\rho_E$ is proportional to $c-1$ in the attractor phase in this model (see \eqref{rho att}). Therefore, what is constrained from the observation is essentially $\rho_E$ at $N=N_{\rm  CMB}$.

%====================================================================================%
\section{Classical versus Stochastic Motion}
\label{Classical versus Stochastic Motion}
%====================================================================================%

In the previous section, we calculated the parameter of the statistical anisotropy
$g_*$ generated in anisotropic inflation. Then it turns out that the model parameter $c$ has to be extremely close to unity to satisfy the observational constraint.  However, the key assumption was that the system had been in the attractor regime till $N=N_{\rm CMB}$ and the background electric energy $\rho_E$ was given by its attractor value $\rho_E^{\rm att}$.

This model has the attractor, because the friction force from the electric field on the inflaton works as a feedback mechanism, $\dot{\varphi}$ stays at smaller value than the case without the electric field (see \eqref{eq: infat}), and the lost part of the inflaton kinetic energy is preserved as the constant electric field. Nevertheless, in the limit $c\to 1$, this feedback virtually vanishes. $\rho_E(t)$ stays constant even in that case, because the coupling is fine-tuned so that $f\propto a^{-2}$ even without the additional friction from $\rho_E$ on $\varphi$. Then there is no reason for $\rho_E$ to approach $\rho_E^{\rm att}$.
%, as long as $\rho_E$ is sufficiently small so that its feedback to $\varphi$ %is negligible. 
In this section, therefore, we revisit the dynamics of the electric field for $N>N_{\rm CMB}$ in the case with $c-1\ll 1$
in order to reconsider the validity of the assumption $\rho_E=\rho_E^{\rm att}$ at $N=N_{\rm CMB}$.

We first consider the classical evolution of $\rho_E$. From \eqref{rho general} we have
\begin{equation} \label{rho E a}
\rho_E(t) = \frac{\rho_E^{\rm att}}{1+\Omega\, a^{-4(c-1)}},
\qquad
\Omega=\frac{\rho_E^{\rm att}}{\rho_E^{\rm ref}}-1,
\end{equation}
where $\rho_E^{\rm ref}\equiv \rho_E(t_{\rm ref})$ is a reference value
at a certain time $t_{\rm ref}$ at which we set $a(t_{\rm ref})=1$.
One can calculate its time variation of $\rho_E$ for 1 e-folding time as
\begin{equation}
\Delta \rho_E^{\rm classical}\equiv \frac{\dd}{H\dd t} \rho_E(t) = \frac{-4\Omega(c-1) a^{-4(c-1)}}{[1+\Omega a^{-4(c-1)}]^2}\rho_E^{\rm att}.
\end{equation}

Next, we consider the stochastic motion of $\rho_E$. From \eqref{f general},
one can see that $f\propto a^{-2}$ is a very good approximation for a tiny $c-1$, even if the system is not in the attractor regime.
Then the fluctuation of the electric field is given by \eqref{delta Ek} with $n=2$. The power spectrum of the electric fluctuation in the super horizon limit is obtained as
\begin{equation}
\mcP_E=\frac{9}{2\pi^2}H^4.
\label{PE}
\end{equation}
where the contributions from the two polarization modes are summed.
It implies that an electric energy density $\delta\rho_E=  \mcP_E/2$
is generated from the vacuum fluctuation every e-folding times on average. Hence we find
\begin{equation}
\Delta \rho_E^{\rm stochastic}\equiv \frac{1}{2}\mcP_E
=\frac{9}{4\pi^2}H^4.
\end{equation}
The ratio between the classical and the stochastic motions is 
\begin{align}
\left|\frac{\Delta \rho_E^{\rm classical}}{\Delta \rho_E^{\rm stochastic}}\right|
&=\frac{3(c-1)^2 |\Omega|\, a^{-4(c-1)}}{\mcP_\zeta^{(0)}[1+\Omega a^{-4(c-1)}]^2}\,,
\notag\\
&= \frac{3|\Omega| (c-1)^2}{(1+\Omega)^2\mcP_\zeta^{(0)}} + \mathcal{O}((c-1)^3)\,,
\notag\\
&\approx 10^{-5}\, \frac{\rho_E(t)|\rho_E(t)-\rho_E^{\rm att}| }{(\rho_E^{\rm att})^2} \left(\frac{c-1}{10^{-7}}\right)^2 \left(\frac{\mcP_\zeta^{(0)}}{2\times10^{-9}}\right)^{-1}\,,
\label{CS ratio}
\end{align}
where $\rho_E^{\rm ref}=\rho_E(t)+\mathcal{O}(c-1)$ is used and $\mathcal{O}((c-1)^3)$ terms are suppressed in the last line.
This result implies that the stochastic motion cannot be ignored for $\rho_E(t)\lesssim10^3 \rho_E^{\rm att}$. 
Even if $\rho_E(t)\simeq \rho_E^{\rm att}$ at some time, the fluctuations $\delta E_k$ are gradually piles up and $\rho_E$ is eventually brought to a much larger value. 
Substituting $\rho_E\simeq 10^3 \rho_E^{att}$ into \eqref{g* expression} with keeping $c-1\approx 10^{-7}$, one obtains $g_* \simeq 10$.
Thus, in the anisotropic inflation model,  it is naturally expected $g_*\simeq 10$,
even though the model parameter $c$ is tuned to be very close to unity. To quantitatively confirm this observation, we shall exploit the stochastic formalism.

%====================================================================================%
\section{Stochastic Formalism of Vector Field}
\label{Stochastic Formalism of Vector Field}
%====================================================================================%

In this section, we develop the stochastic formalism of the electric field in the anisotropic inflaton model.
We adopt not the gauge field $A_i$ but the electric field $E_i$ as a main variable, since $E_i$ stays constant
on super-horizon scales in anisotropic inflation and can be treated in the similar way to a massless scalar field.
To go to the stochastic picture, we divide the electric field into the IR\ part and the UV  part,
\begin{equation}
E_i(t,\bm{x})= E_i^{\rm IR}(t,\bm{x}) + E_i^{\rm UV}(t,\bm{x}),
\label{UVIR dec}
\end{equation}
with
\begin{align}
E_i^{\rm IR}(t,\bm{x}) &\equiv \int \frac{\dd^3 k}{(2\pi)^3} e^{-i\bm{k}\cdot\bm{x}}\,W(t,k)
E_i(t,\bm{k}),
\\
E_i^{\rm UV}(t,\bm{x}) &\equiv \int \frac{\dd^3 k}{(2\pi)^3} e^{-i\bm{k}\cdot\bm{x}}\left[1-W(t,k)\right]
E_i(t,\bm{k}),
\end{align}
where $E_i(t,\bm{k})\equiv \int \dd^3 x e^{i\bm{k}\cdot\bm{x}} E_i(t,\bm{x})$ is the normal Fourier transformation of $E_i(t,\bm{x})$ and we introduce a window function,
\begin{equation}
W(t,k)= \Theta(\epsilon a H-k).
\end{equation}
Here $\Theta(x)$ is the Heaviside function and $\epsilon\ll1 $ is a small constant parameter. $E_i^{\rm IR}(t,\bm{x})$ contains only the contributions from the mode functions of the electric field on much larger scales than the horizon, $k< \epsilon a H$.
We also define the IR and UV part of its conjugate momentum $\Pi_i(t,\bm{x})\equiv \dot{E}_i(t,\bm{x})$ as
\begin{align}
\Pi_i^{\rm IR}(t,\bm{x}) &\equiv \int \frac{\dd^3 k}{(2\pi)^3} e^{-i\bm{k}\cdot\bm{x}}\, W(t,k)
\dot{E}_i(t,\bm{k}),
\\
\Pi_i^{\rm UV}(t,\bm{x}) &\equiv \int \frac{\dd^3 k}{(2\pi)^3} e^{-i\bm{k}\cdot\bm{x}}\left[1-W(t,k)\right]
\dot{E}_i(t,\bm{k}).
\end{align}

Now we consider the evolution of $E_i^{\rm IR}$.
The EoM for the original electric field in coordinate space is given by
\begin{equation}
\ddot{E}_i(t,\bm{x})+5H \dot{E}_i(t,\bm{x})-a^{-2}\partial_j^2 E_i(t,\bm{x})
+\left[6H^2+\frac{\ddot{f}+H\dot{f}}{f}-2\frac{\dot{f}^2}{f^2}\right]E_i(t,\bm{x})
=0.
\label{EEoM t}
\end{equation}
where we ignored the slow-roll corrections and the spatial derivative of $f$. The gauge kinetic function $f(\bar{\varphi}(t))$ should be obtained by simultaneously solving the EoM for $\bar{\varphi}(t)$. However, since we are interested only in the electric field, we derive an effective EoM only for $E_i^{\rm IR}$
by using the classical background solution, $f\propto a^{-2}\sqrt{1+\Omega a^{-4(c-1)}}$, and making an approximation. From the background solution, one finds
\begin{equation}
6H^2+\frac{\ddot{f}+H\dot{f}}{f}-2\frac{\dot{f}^2}{f^2}
\simeq -\frac{10\Omega H^2}{1+\Omega}(c-1)
\simeq 10H^2\left[\frac{\bm{E}^2}{2\rho_E^{\rm att}}-1\right](c-1)\,,
\label{E mass}
\end{equation}
where the $\mathcal{O}((c-1)^2)$ terms are ignored and $\rho_E^{\rm ref}\simeq \rho_E(t)=\bm{E}^2(t)/2$ is used.
%Note that the electric field in the above equation is the background one %$E_i(t)\equiv -f\dot{A}_{\rm BG}/a$ and it is different from the local one %$E_i(t,\bm{x})$ in \eqref{EEoM t}. 
Here, we approximate the background electric field $E_i(t)$ by the IR mode $E_i^{\rm IR}(t,\bm{x})$
and obtain 
\begin{equation}
\ddot{E}_i(t,\bm{x})+5H \dot{E}_i(t,\bm{x})-a^{-2}\partial_j^2 E_i(t,\bm{x})
+10(c-1)H^2\left[\frac{\bm{E}^2_{\rm IR}(t,\bm{x})}{E_{\rm att}^2}-1\right]E_i(t,\bm{x})
=0,
\label{EEoM}
\end{equation}
where $E_{\rm att}\equiv \sqrt{2\rho_E^{\rm att}}.$
With these ingredients, it can be shown that the EoM for $E_i^{\rm IR}$ is derived as
(see appendix~\ref{Derivation of the Langevin Equation} for detailed derivation)
\begin{equation}
\dot{E}_i^{\rm IR}
+2(c-1)H\left[\frac{\bm{E}^2_{\rm IR}}{E_{\rm att}^2}-1\right]E_i^{\rm IR}
=\xi_i^E(t),
\label{Langevin eq}
\end{equation}
which is a Langevin equation.
Here, $\xi_i^E$ representing a stochastic term satisfies 
(see \eqref{xixi})
\begin{equation}
\left\langle \xi_i^E(t)\right\rangle=0, \qquad
\left\langle \xi_i^E(t) \xi_j^E(t')\right\rangle =
\frac{1}{3}H\mcP_E(k_c)\delta_{ij}\delta(t-t')\simeq \frac{3H^5}{2\pi^2}\delta_{ij}\delta(t-t'),
\end{equation}
where $k_c(t)=\epsilon a(t) H$ and \eqref{PE} is used.
Compare to the slow-roll equation, $5H\dot{\bm{E}}_{\rm IR}+\partial_{\bm E} V_{\rm eff}(\bm{E}_{\rm IR})=\bm{\xi}_E$, one finds a Mexican-hat type effective potential of $E_i^{\rm IR}$
\begin{equation}
V_{\rm eff}(\bm{E}_{\rm IR})=\frac{20\pi^2}{3H^2}\mcP_\zeta^{(0)}
\, \bm{E}^2_{\rm IR}
\left[\bm{E}^2_{\rm IR}- 2E_{\rm att}^2\right],
\end{equation}
and the potential minimum is located at $\bm{E}^2_{\rm IR}=E_{\rm att}^2$ as expected.
The curvature of the potential around the minimum is small, $V''_{\rm eff}(E_{\rm att})=20H^2(c-1)$, and the electric field is not stabilized there.
On the other hand, the quartic potential does not depend on $c-1$, and the potential curvature grows as $V''_{\rm eff}(\bm{E}_{\rm IR})\simeq 80\pi^2 \mcP_\zeta^{(0)}\bm{E}^2_{\rm IR}/H^2\approx 10^{-6}\bm{E}^2_{\rm IR}/H^2$ for $\bm{E}_{\rm IR}^2\gg \rho_E^{\rm att}$. Hence it is unlikely that $\bm{E}^2_{\rm IR}$ becomes much larger than $10^6 H^4$. 

%
%///////////////////////////////////////////////////////////////////////////////////%
\begin{figure}[tbp]
  \begin{center}
  \includegraphics[width=110mm]{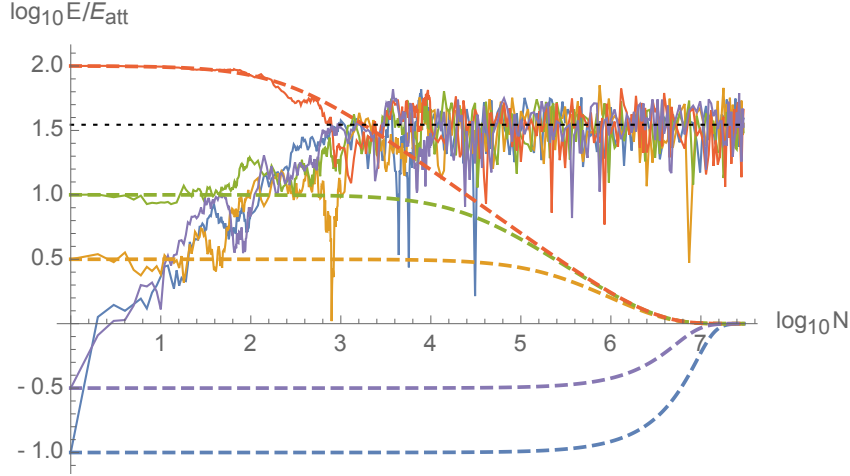}
  \end{center}
  \caption
 {We compare the classical background solution without the fluctuation (dashed lines) and the numerical solution of the Langevin equation \eqref{Langevin eq} (solid lines). We set the model parameter $c-1=10^{-7}$ and the initial values $|\bm{E}_{\rm IR}|/E_{\rm att}=10^2$ (red), $10$ (green), $10^{1/2}$ (yellow), $10^{-1/2}$ (purple) and $10^{-1}$ (blue).
Although all the classical solutions merge into the attractor $|\bm{E}_{\rm IR}|=E_{\rm att}$ for the time scale $N\simeq (c-1)^{-1}$, the solutions of the Langevin equation stay at around $|\bm{E}_{\rm IR}|\simeq 10^{3/2}E_{\rm att}$ due to the stochastic effect given by the super-horizon fluctuations of the electric field. The black dashed line represents
the expected value of $|\bm{E}_{\rm IR}|/E_{\rm att}$ in the equilibrium state obtained in \eqref{E expected value}.}
 \label{langevin_fig}
\end{figure}
%///////////////////////////////////////////////////////////////////////////////////%
%
In Fig.~\ref{langevin_fig}, we show the results of the numerical calculations of \eqref{Langevin eq} with and without the noise term $\xi_i^E$. 
The solutions without the noise term agree with the background solutions \eqref{rho general}. To numerically solve \eqref{Langevin eq}, we reduce them into three coupled difference equations with one e-fold step size and take into account $\xi_i^E$ as the Gaussian distributed random number with the variance $3H^5/2\pi^2$ for each step.
For illustrative purpose, in Fig.~\ref{langevin_fig}, we do not plot all of the $\mathcal{O}(10^7)$ points, while we show only up to $\mathcal{O}(10^2)$ points in the each bin of $\Delta \ln_{10}N=1$.
%\textcolor{red}{In order to lighten the image processing of the figure, %we omitted to plot all the data of step at late time and arranged to reflect %the same number of plot as that in early time.}
 In Fig.~\ref{langevin_fig}, one can see that the numerical solutions of the Langevin equation do not approach the attractor but they
typically stay at much larger values, contrary to the classical background solutions going straight to the attractor solution.
This result visibly implies that the electric field averaged over a certain scale (e.g. the scale of our observable universe) is typically much stronger
than the attractor value in anisotropic inflation despite  small $c-1$.
It should be also stressed that all the numerical stochastic solutions reach the equilibrium at $N\sim 10^3$ which is much earlier than
the classical solutions reach the attractor at $N\sim 10^7$.

%====================================================================================%
\section{Probability Distribution of Anisotropy}
\label{Probability Distribution of Anisotropy}
%====================================================================================%

In this section, we calculate the probability distribution function (PDF) of the anisotropic parameter $g_*$ in the anisotropic inflation model.
To this end, we focus on the norm of the electric field,  $E_{\rm IR}\equiv |\bm{E}^{\rm IR}|$, because the electric energy $\rho_E$ determining $g_*$ depends only on the norm.
PDF can be obtained by solving the Fokker-Planck equation. 
The Fokker-Planck equation %associated with the Langevin equation \eqref{Langevin eq} 
for the norm of the IR mode electric field is (see appendix \ref{Derivation of the Fokker-Planck Equation} for  derivation)
\begin{equation}
\frac{\partial P(\mathcal{E},\mathcal{N})}{\partial \mathcal{N}}
=
\frac{\partial}{\partial \mathcal{E}}\left[ \frac{\partial U}{\partial \mathcal{E}}
%\left(16\mcP_\zeta^{(0)}\mathcal{E}^3-2(c-1)\mathcal{E}-\frac{2}{\mathcal{E}} \right)
\,P(\mathcal{E},\mathcal{N}) \right]+
\frac{\partial^2}{\partial \mathcal{E}^2}P(\mathcal{E},\mathcal{N}),
\label{FP eq}
\end{equation}
with the potential 
%(the meaning of the logarithmic term will be explained soon),
%
\begin{equation}
U(\mathcal{E})=-(c-1)\mathcal{E}^2+4\mcP_\zeta^{(0)}\mathcal{E}^4-2\ln\mathcal{E}.
\end{equation}
Here we have introduced dimensionless variables,
\begin{equation}
\mathcal{E} \equiv \sqrt{\frac{c-1}{8\mcP_\zeta^{(0)}}}\frac{E_{\rm IR}}{E_{\rm att}}\approx 2.5\frac{E_{\rm IR}}{E_{\rm att}}\left(\frac{c-1}{10^{-7}}\right)^{\frac{1}{2}},\qquad \dd\mathcal{N}\equiv H\dd t.
\end{equation}
The Fokker-Planck equation \eqref{FP eq} describes the time evolution of the probability distribution of $\mathcal{E}$ at an e-folds $\mathcal{N}$.
Although it is generally difficult to find its solution,
one can easily obtain the static equilibrium solution as
\begin{equation}
P_{\rm eq}(\mathcal{E})=\mathrm{N}^{-1}\exp\left[-U(\mathcal{E})\right],
\label{Peq}
\end{equation}
where $\mathrm{N}=\int^\infty_0\dd \mathcal{E} \exp(-U)$ is the normalization factor.

With this equilibrium probability distribution for $\mathcal{E}$, we can calculate various expectation values associated with $E_{\rm IR}$.
In the case with $c-1=10^{-7}$, the expectation values of $E_{\rm IR}/ E_{\rm att}$ and $\rho_E^{\rm IR}/\rho_E^{\rm att}$ are
\begin{align}
&\frac{\langle E_{\rm IR}\rangle}{E_{\rm att}}=\sqrt{\frac{8\mcP_\zeta^{(0)}}{c-1}}\langle\mathcal{E}\rangle
=\sqrt{\frac{8\mcP_\zeta^{(0)}}{c-1}}
\int_0^\infty \dd \mathcal{E}\, \mathcal{E} P_{\rm eq}(\mathcal{E})\,
\approx 35,
\label{E expected value}
\\
&\frac{\langle\rho_E^{\rm IR}\rangle}{\rho_E^{\rm att}}=\frac{\langle E_{\rm IR}^2\rangle}{E_{\rm att}^2}
=\frac{8\mcP_\zeta^{(0)}}{c-1}
\int_0^\infty \dd \mathcal{E}\, \mathcal{E}^2 P_{\rm eq}(\mathcal{E})
\approx 1400.
\label{rho expected value}
\end{align}
\eqref{E expected value} is compared to the numerical solutions of the Langevin equation \eqref{Langevin eq} in Fig.~\ref{langevin_fig} and they show a great agreement.
Substituting \eqref{rho expected value} into \eqref{g* expression}, we find that the averaged value of the parameter of the statistical anisotropy is
\begin{equation}
\langle g_* \rangle \simeq  10,
\end{equation}
which is far beyond the observational upper bound, $|g_*| \lesssim 10^{-2}$.

%
%///////////////////////////////////////////////////////////////////////////////////%
\begin{figure}[tbp]
  \begin{center}
  \includegraphics[width=110mm]{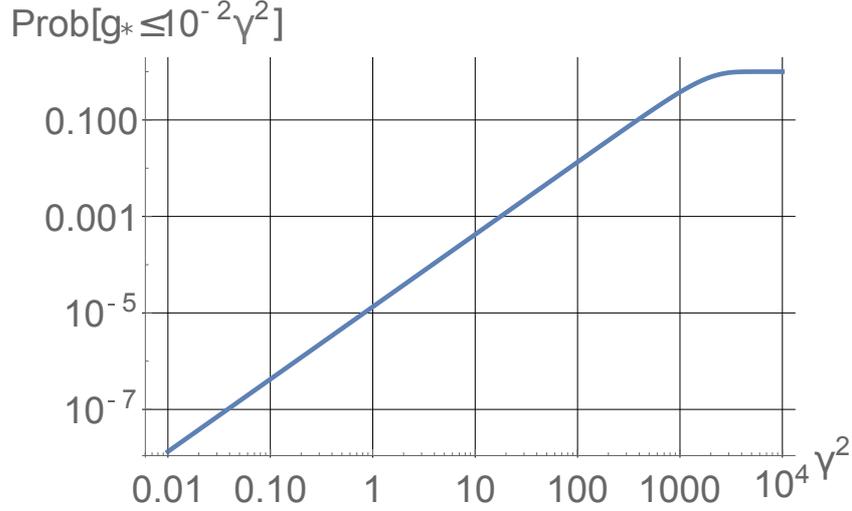}
  \end{center}
  \caption
 {We show the probability that $g_*$ does not exceed $0.01 \gamma^2$ as a function of $\gamma^2$. The model parameter is set as $c-1=10^{-7}$. The plotted line is well approximated by $1.3\times 10^{-5}\gamma^3$
for $\gamma^2<10^3$. The CMB observation puts the constraint $g_*\lesssim10^{-2}$,
and the probability to satisfy it is around $10^{-5}$.}
 \label{Prob1}
\end{figure}
%///////////////////////////////////////////////////////////////////////////////////%
%
It is interesting to see the probability that $g_*$ satisfies the observational constraint.
The probability that $E_{\rm IR}$ does not exceed  $E_{\rm att}$ multiplied by a constant factor $\gamma$ in the equilibrium state is given by
\begin{align}
{\rm Prob}\left(E_{\rm IR} \le \gamma E_{\rm att}\right)
=
\int_0^{\gamma \sqrt{(c-1)/8\mcP_\zeta^{(0)}}} \dd \mathcal{E} P_{\rm eq}(\mathcal{E})\,,
\label{Prob gamma}
%\notag\\
%&=\mathrm{N}^{-1}
%\int_0^{\gamma \sqrt{(c-1)/8\mcP_\zeta^{(0)}}} \dd \mathcal{E}\, \mathcal{E}^2 %\exp\left[(c-1)\mathcal{E}^2-4\mcP_\zeta^{(0)}\mathcal{E}^4\right].
\end{align}
This ${\rm Prob}\left(E_{\rm IR} \le \gamma E_{\rm att}\right)$ 
can be identified as the probability of $g_*\le 10^{-2}\gamma^2$, provided that the electric field reached the equilibrium until $N=N_{\rm CMB}$ and the electric field norm $|\bm{E}(t,\bm{x})|$ averaged over our observable universe took a value distributed as \eqref{Peq} at $N=N_{\rm CMB}$.
Numerically integrating \eqref{Prob gamma}, we obtain the probability for $\gamma=1$ and $c-1=10^{-7}$ as
\begin{equation}
{\rm Prob}(E_{\rm IR}\le E_{\rm att})\approx 1.3\times 10^{-5}.
\qquad \left(c-1=10^{-7}\right)
\end{equation}
Therefore, the probability that $g_*$ is not larger than $10^{-2}$ is
$0.0013\%$. In other words, the prediction of the anisotropic inflation model with $c-1=10^{-7}$ is excluded with $99.9987\%$  by the observational constraint on $g_*$, if the gauge field is in the equilibrium state before the horizon crossing of the CMB modes.
In Fig.\ref{Prob1}, we show \eqref{Prob gamma} with respect to $\gamma^2$.

\begin{figure}[tbp]
  \begin{center}
  \includegraphics[width=110mm]{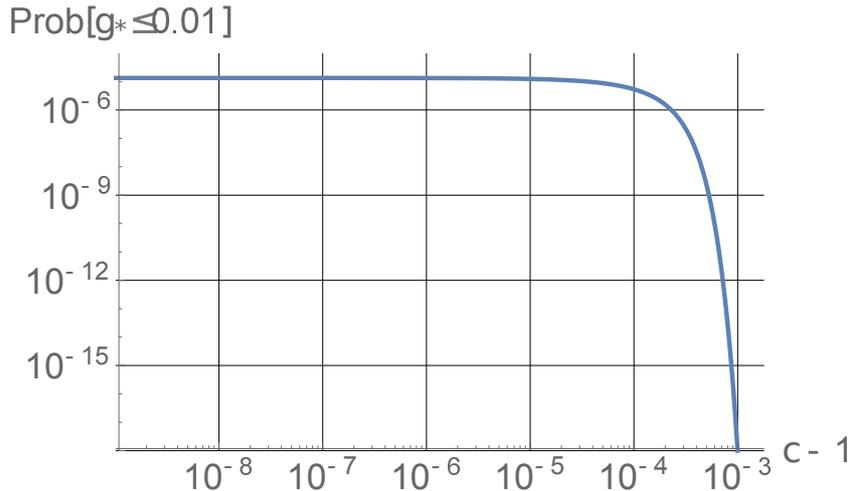}
  \end{center}
  \caption
 {The probability that the predicted $g_*$ is small enough to be compatible with the CMB observation, namely $g_*\le 10^{-2}$, is shown as a function of $c-1$. As $c-1$ decreases, the probability increases. However, the increase
of the probability saturates at $c-1\simeq 10^{-5}$. 
The probability is always $\approx 1.3\times 10^{-5}$ for $c-1\lesssim 10^{-5}$.}
 \label{Prob2}
\end{figure}

Using \eqref{Prob gamma}, we can also find the  probability of $g_* \le 0.01$ as a function of $c$.
Note that as $c$ changes, $E_{\rm att}\propto \sqrt{c-1}$ also changes, but $E_{\rm IR}$ should be smaller than the constant value of $E_{\rm att}$ with $c-1=10^{-7}$. Thus the probability which we consider is given by  
${\rm Prob}(E_{\rm IR}\le \sqrt{10^{-7}/(c-1)}E_{\rm att})$. 
It is shown in Fig.\ref{Prob2}.
%
%///////////////////////////////////////////////////////////////////////////////////%
%///////////////////////////////////////////////////////////////////////////////////%
%
One can see that the probability to satisfy the CMB constraint is the same as the case $c-1=10^{-7}$ as long as $c-1\lesssim 10^{-5}$. The probability further drops for a $c-1\gtrsim 10^{-4}$,
because the classical motion wins the stochastic motion 
(see \eqref{CS ratio}),
the electric field approaches the attractor value, but
it is too large.

Note that if $c-1\gtrsim 10^{-3}$,
the $c-1$ expansion used in \eqref{E mass} quickly breaks down for a typical energy density $\rho_E(t)\simeq 10^3 \rho_E^{\rm att}$ and hence we plot the probability up to $c-1=10^{-3}$ in Fig.~\ref{Prob2}.

%====================================================================================%
\section{Summary and Discussion}
\label{Summary and Discussion}
%====================================================================================%

The statistical anisotropy of the curvature perturbation is constrained to be less than a few percent by the CMB observations. It was discussed that 
%the energy fraction of background electric field in the  attractor solution
the model parameter $c$
was required to be extremely close to unity in the framework of anisotropic inflation.
However, it makes the curvature of the effective potential of the gauge field around the attractor solution almost flat and hence quantum kicks to the background is expected to be relevant. This situation is analogous 
to a scalar field case where stochastic effect becomes significant.
Therefore in this paper, we developed the stochastic formalism of a vector field in the context of anisotropic inflation and estimated the probability that the parameter of statistical anisotropy $g_*$ satisfies the current observational constraint.

In Sec.\ref{Classical versus Stochastic Motion}, we presented a naive estimation of the parameter region where the stochastic force to the background vector field cannot be ignored compared with its classical force.
%There we found that stochastic motion becomes much significant and background gauge field is easily brought to be a large value compared with an attractor value.
In Sec.\ref{Stochastic Formalism of Vector Field}, we derived the Langevin equation of the electric field and numerically showed that the solutions stay at much larger values than that of attractor.
%before the horizon crossing of CMB scales.
% More precisely, we derived a Langevin equation of vector field and showed an evolution of background vector field induced by the quantum kicks to the background, and analyzed the probability distribution function of parameter of anisotropy by solving an associated Fokker-Planck equation.
 Finally, in Sec.\ref{Probability Distribution of Anisotropy} we solved its Fokker-Planck equation and revealed that the probability that the generated statistical anisotropy is compatible with the current upper bound cannot be larger  than 0.0013\% irrespective of the model parameter of anisotropic inflation. Equivalently, the anisotropic inflation scenario is excluded with around 99.999\% by the current observational bound.
It should be also emphasized that we obtained the same quantitative results of the expected anisotropy $g_* \simeq 10$ (i.e. $\rho^{\text{IR}}_E \simeq 10^3\rho^{\text{att}}_E$) in the three different ways in Sec.~\ref{Classical versus Stochastic Motion}, \ref{Stochastic Formalism of Vector Field} and \ref{Probability Distribution of Anisotropy}.
 %Remarkably, our result is consistent with the previous study without stochastic method.
 %As a consequence, the probability that the parameter of anisotropy is not larger than current upper bound is around 0.001\% irrespective of the model parameter of anisotropic inflation, which means that an anisotropic inflation is excluded with around 99.999\% under the current observations.

One can discuss a way to alleviate the above problem.
Aside from the model parameter ~$c-1$, 
the ratio of the classical and the stochastic motion depends on
$\mathcal{P}_\zeta^{(0)}$ (see \eqref{CS ratio}).
It should be noted that we used $\mathcal{P}_\zeta^{(0)} \approx 2.2\times 10^{-9}$ through our derivation. 
One can seek a possibility to avoid the small probability,
by considering another mechanism of generating curvature perturbation such as the curvaton mechanism~\cite{Enqvist:2001zp,Lyth:2001nq,Moroi:2002rd}.
The observational constraint on $g_*$ can be also avoided, when the gauge field is coupled not to the inflaton but to a spectator field. In this case, we don't need a fine-tuning of a model parameter and a small anisotropy allowed by the current observation can be naturally generated due to the gravity coupling to the inflaton.
 Intriguingly, at the same time a sizable statistical anisotropy of primordial gravitational waves can be provided~\cite{FOTY}.
Meanwhile, it would be interesting to explore the stochastic approach to the other scenarios where a vector field plays an important role during inflation.
 We leave these issues in our future work.

%====================================================================================%
\section{Acknowledgement}
%====================================================================================%

 We would like to thank T. Tanaka, Y. Tada and J. Tokuda for fruitful discussions and comments.
 The part of this work is supported by JSPS KAKENHI Grant Number 15J01345 and the Grant-in-Aid for JSPS Fellows No.~29-9103.

\appendix

%====================================================================================%
\section{Derivation of the Langevin Equation}
\label{Derivation of the Langevin Equation}
%====================================================================================%

Here we derive the Langevin equation \eqref{Langevin eq}.
From the definitions of $E_i^{\rm IR}(t,\bm{x}), E_i^{\rm UV}(t,\bm{x})$,
$\Pi_i^{\rm IR}(t,\bm{x})$ and $\Pi_i^{\rm IR}(t,\bm{x})$,
one finds
\begin{equation}
\Pi_i^{\rm IR}(t,\bm{x}) +\Pi_i^{\rm UV}(t,\bm{x})
=\dot{E}_i^{\rm IR}(t,\bm{x}) +\dot{E}_i^{\rm UV}(t,\bm{x}).
\end{equation}
It yields an equation only for the IR mode,
\begin{equation}
\dot{E}_i^{\rm IR}(t,\bm{x})-\Pi_i^{\rm IR}(t,\bm{x})
=\Pi_i^{\rm UV}(t,\bm{x})- \dot{E}_i^{\rm UV}(t,\bm{x})
= \int \frac{\dd^3 k}{(2\pi)^3} e^{-i\bm{k}\cdot\bm{x}}\,\dot{W}(t,k)
E_i(t,\bm{k}),
\label{EPPE}
\end{equation}
where the definition of the UV modes are used. 
Substituting
\begin{equation}
\dot{W}(t,k)=\epsilon aH^2\delta(\epsilon aH-k),
\end{equation}
into \eqref{EPPE}, we obtain
\begin{equation}
\dot{E}_i^{\rm IR}-\Pi_i^{\rm IR}=\xi_i^E\equiv 
%\Pi_i^{\rm UV}-\dot{E}_i^{\rm UV} =
\epsilon a H^2\int \frac{\dd^3 k}{(2\pi)^3} e^{-i\bm{k}\cdot\bm{x}}\,\delta(\epsilon a H-k)
E_i(t,\bm{k}),
\label{EIR EoM}
\end{equation}

Next, rewriting \eqref{EEoM} in terms of $E_i^{\rm IR}, E_i^{\rm UV}$,
$\Pi_i^{\rm IR}$ and $\Pi_i^{\rm IR}$, one finds
\begin{align}
\dot{\Pi}_i^{\rm IR}+5H\Pi_i^{\rm IR}&-a^{-2}\partial_j^2 E_i^{\rm IR}
+10(c-1)H^2\left[\frac{\bm{E}^2_{\rm IR}}{E_{\rm att}^2}-1\right]E_i^{\rm IR}
\notag\\&=-\dot{\Pi}_i^{\rm UV}-5H\Pi_i^{\rm UV}+a^{-2}\partial_j^2 E_i^{\rm UV}
-10(c-1)H^2\left[\frac{\bm{E}^2_{\rm IR}}{E_{\rm att}^2}-1\right]E_i^{\rm UV}
\notag\\&=\int\frac{\dd^3 k}{(2\pi)^3} e^{-i\bm{k}\cdot\bm{x}}\,\dot{W}(t,k)
\dot{E}_i(t,\bm{k})=\xi_i^\pi,
\label{PIIR EoM}
\end{align}
where we have used the Fourier transformed \eqref{EEoM}.
Neglecting the slow-roll suppressed term $\dot{\Pi}_i^{\rm IR}$, the gradient term $a^{-2}\partial_j^2 E_i^{\rm IR}$ and the small noise term  $\xi_i^\pi$, we obtain\footnote{For a sufficiently small $\epsilon$, the gradient term $a^{-2}\partial_j^2 E_i^{\rm IR}$ and $\xi_i^\pi$ in \eqref{PIIR EoM} can be neglected. As one can check with \eqref{dA solution}, $\dot{E}_i(t,\bm{k})$ quickly decays on super-horizon scales. }

\begin{equation}
\Pi_i^{\rm IR}\simeq -2(c-1)H\left[\frac{\bm{E}^2_{\rm IR}}{E_{\rm att}^2}-1\right]E_i^{\rm IR}.
\end{equation}
Plugging it into \eqref{EIR EoM}, we obtain the Langevin equation \eqref{Langevin eq}.
Note that the expectation values of the noise term
$\xi_i^E$ is vanishing because $E_i(t,\bm{k})$ is a perturbation,
while its variance is finite.
The correlation function  of the noise term is calculated as
\begin{align}
\left\langle \xi_i^E(t) \xi_j^E(t')\right\rangle 
&=
k_c(t)k_c(t')H^2\int \frac{\dd^3 k\dd^3 k'}{(2\pi)^6}
\delta(k_c(t)-k)\delta(k_c(t')-k')
\, \langle E_i(t,\bm{k}) E_j(t',\bm{k}')\rangle
\notag\\
&=
k_c^2(t)H^2\int \frac{\dd^3 k}{(2\pi)^3}
\delta(k_c(t)-k) \frac{\delta(t-t')}{k_c(t)H}
\left(\delta_{ij}-\frac{k_ik_j}{k^2}\right)\frac{\pi^2}{k^3}\mcP_E(t,k),
\notag\\
&=
\frac{1}{2} H\mcP_E(t,k)\delta(t-t')\int \frac{\dd \Omega}{4\pi}
\left(\delta_{ij}-\frac{k_ik_j}{k^2}\right),
\notag\\
&=\frac{1}{3}H\mcP_E(k_c)\delta_{ij}\delta(t-t'),
\label{xixi}
\end{align}
Note that $\mcP_E$ given in \eqref{PE} includes the contributions from the two polarization modes and $\langle E_i(t,\bm{k}) E_j(t',\bm{k}')\rangle=
\left(\delta_{ij}-k^{-2} k_ik_j\right)2\pi^2 k^{-3}\mcP_E/2$.

%====================================================================================%
\section{Derivation of the Fokker-Planck Equation}
\label{Derivation of the Fokker-Planck Equation}
%====================================================================================%

Here we derive the Fokker-Planck equation \eqref{FP eq}.
The Langevin equation \eqref{Langevin eq}
can be recast into a dimensionless form,
\begin{equation}
\frac{\dd}{\dd \mathcal{N}}X_i
=\alpha X_i+\beta\bm{X}^2 X_i+\xi_i^X,
\end{equation}
with
\begin{equation}
\alpha\equiv 2(c-1),\quad
\beta\equiv -16\mcP_\zeta^{(0)},
\quad
X_i\equiv \sqrt{\frac{c-1}{8\mcP_\zeta^{(0)}}}\frac{E_i}{E_{\rm att}},
\quad \xi_i^X\equiv 
\sqrt{\frac{c-1}{8\mcP_\zeta^{(0)}}}\frac{\xi_i^E}{H^2 E_{\rm att}}.
\end{equation}
The correlation of this noise term is given by
\begin{equation}
\langle \xi_i^X(\mathcal{N}) \xi_i^X(\mathcal{N}')\rangle=2\delta_{ij}\delta(\mathcal{N}-\mathcal{N}').
\end{equation}
The corresponding Fokker-Planck equation for the PDF of the all the components of the vector $\bm{X}=\{X_1, X_2, X_3\}$ is known as~\cite{Risken:1996tex}
\begin{align}
\frac{\partial}{\partial \mathcal{N}}
P\left(\bm{X},\mathcal{N}\right)
&=-(3\alpha+5\beta \bm{X}^2)P\left(\bm{X},\mathcal{N}\right)
-(\alpha X_i+\beta \bm{X}^2 X_i)
\frac{\partial}{\partial X_i} P\left(\bm{X},\mathcal{N}\right)
\notag\\
&\quad+\frac{\partial^2}{\partial X_i^2} P\left(\bm{X},\mathcal{N}\right).
\label{FP X}
\end{align}
This equation describes the evolution of the PDF which gives the expectation value of an arbitrary function of the three components of the vector,
\begin{equation}
\langle f(X_1,X_2,X_3) \rangle
= \int \dd X_1\, \dd X_2\, \dd X_3\, 
f(X_1,X_2,X_3)
P(\bm{X},\mathcal{N}).
\end{equation}
However, we are interested only in the norm of the vector $|\bm{X}|$
because $g_*$ is proportional to the energy density of the electric
field $\rho_E=|\bm{E}|^2/2$.
In order to reduce the full PDF $P\left(\bm{X},\mathcal{N}\right)$
into the PDF of only the norm, we change the variable $\{X_1,X_2,X_3\}$ into the polar coordinate $\{R,\theta,\varphi\}$ and then
the probability for the infinitesimal volume is 
written as
\begin{equation}
P(\{X_1,X_2,X_3\},\mathcal{N})\dd X_1\, \dd X_2\, \dd X_3
=\tilde{P}(\{R,\theta,\varphi\},\mathcal{N})R^2\sin\theta\,\dd R\, \dd \theta\, \dd \varphi,
\label{polar C}
\end{equation}
where the radial coordinate $R$ corresponds to the norm of the electric field.
Performing the angular integral in the both sides of \eqref{FP X}, one finds
\begin{align}
\frac{\partial}{\partial \mathcal{N}}
\bar{P}\left(R,\mathcal{N}\right)
&=-(2R^{-2}-\alpha+3\beta R^2)\bar{P}\left(R,\mathcal{N}\right)
-(2R^{-1}+\alpha R +\beta R^3)\frac{\partial}{\partial R} \bar{P}\left(R,\mathcal{N}\right)
\notag\\
&\quad+ \frac{\partial^2}{\partial R^2} \bar{P}\left(R,\mathcal{N}\right)
+\int\dd \Omega \left[
\frac{1}{\sin\theta}\frac{\partial}{\partial\theta}\left(\sin\theta
\frac{\partial\tilde{P}}{\partial\theta}\right)+\frac{1}{\sin^2\theta}\frac{\partial^2\tilde{P}}{\partial\varphi^2}\right].
\label{angular rem}
\end{align}
Here we introduce the PDF only for the norm,
\begin{equation}
\bar{P}\left(R,\mathcal{N}\right)=R^2\int\dd \Omega\,
\tilde{P}(\{R,\theta,\varphi\},\mathcal{N}),
\end{equation}
where the Jacobian factor $R^2$ is included.
The last term in \eqref{angular rem} vanishes when
\begin{equation}
\left[\sin\theta \frac{\partial\tilde{P}}{\partial\theta}\right]_{\theta=\pi}
-
\left[\sin\theta \frac{\partial\tilde{P}}{\partial\theta}\right]_{\theta=0}=0,
\qquad
\left[\frac{\partial\tilde{P}}{\partial\varphi}\right]_{\varphi=2\pi}
-\left[\frac{\partial\tilde{P}}{\partial\varphi}\right]_{\varphi=0}=0.
\end{equation}
If $\partial\tilde{P}/\partial\theta$ is not singular at $\theta=0,\pi$
and $\partial \tilde{P}/\partial\varphi$ is periodic with respective to $\varphi=[0,2\pi]$,
the above conditions are satisfied.
We assume they are satisfied.
In that case, the Fokker-Planck equation reads
\begin{align}
\frac{\partial}{\partial \mathcal{N}}
\bar{P}\left(R,\mathcal{N}\right)
&=\frac{\partial}{\partial R}\left[ \frac{\partial U}{\partial R}
\,\bar{P}(R,\mathcal{N}) \right]+
\frac{\partial^2}{\partial R^2}\bar{P}(R,\mathcal{N}),
\end{align}
with the mexican hat type potential with a logarithmic term,
\begin{equation}
U(R)=-2\ln R-\frac{\alpha}{2}R^2-\frac{\beta}{4}R^4.
\end{equation}
By renaming $\bar{P}\to P$ and $R\to \mathcal{E}$,
we obtain \eqref{FP eq}.
One may be puzzled by the logarithmic term in the above potential. 
It reflects that we have reduced the PDF of the three dimensional vector
into that of scalar (i.e. the norm of the vector) and the Jacobian factor
$R^2$ seen in \eqref{polar C} should be restored in the calculation of probabilities
with $\bar{P}(R,\mathcal{N})$.
Indeed, one can explicitly see that the Jacobian factor appears in the case of the static equilibrium probability \eqref{Peq},
\begin{equation}
P_{\rm eq}(R)\dd R= \mathrm{N}^{-1}\exp\left[-U(R)\right]\dd R=\mathrm{N}^{-1}
\exp\left[\frac{\alpha}{2}R^2+\frac{\beta}{4}R^4\right] R^2\dd R,
\end{equation}
where only the mass term and the quartic interaction term remain in the exponent.

\end{document}